\documentclass[
prx,
aps,
twocolumn,
10pt
]{revtex4-1}

\usepackage{graphicx}

\usepackage{amsmath}

\usepackage{amssymb}

\usepackage{blkarray}

\usepackage{bbold}

\usepackage{braket}

\begin{document}
\title{Breathing kagome XY quantum magnet with four-site ring exchange}

\author{Niklas Casper}
\email{n.casper@tu-bs.de}
\affiliation{Institute for Theoretical Physics, Technical University
Braunschweig, D-38106 Braunschweig, Germany}

\author{Wolfram Brenig}
\email{w.brenig@tu-bs.de}
\affiliation{Institute for Theoretical Physics, Technical University
Braunschweig, D-38106 Braunschweig, Germany}

\date{\today}


\begin{abstract}We study the impact of trimerization (breathing) of the
nearest-neighbor (NN) exchange on the planar XY spin-$1/2$ ferromagnet on the
kagome lattice, including additional four-site ring exchange. For uniform NN
exchange, this model has previously been shown to transit from a long-range
ordered ferromagnet into a $\mathbb{Z}_{2}$ quantum spin liquid by virtue
of the ring exchange. Using quantum Monte-Carlo calculations, based on the
stochastic series expansion, we present results for the spin stiffness, the
quantum phase diagram, the longitudinal static, as well as the transverse
dynamic structure factor. Our results corroborate 3D XY universality also at
finite trimerization and suggest a simple continuation of the quantum critical
point of the uniform case into a line in terms of rescaled exchange
parameters. Moreover, at any trimerization and in the ordered phase the
elementary excitations can be understood very well in terms of linear spin wave
theory, while beyond the critical line, in the spin liquid phase we find
signatures of spinon continua.\end{abstract}

\maketitle


\section{Introduction}

Quantum spin liquids (QSL) constitute intriguing forms of magnetic matter, which
are attracting great interest since several decades by now \cite{Anderson1973,
Fazekas1974, Misguich2005, Balents2010, Savary2017, Knolle2019}. QSLs are
characterized by the absence of local magnetic order parameters, even at zero
temperature, they show fractionalized excitations, long-ranged topological
entanglement, and quantum orders beyond Landau's paradigm of symmetries and
spontaneous symmetry breaking. Proposals for putative QSLs mostly rest on an
underlying gauge structure \cite{Wen2002}, e.g. $U(1)$ or $\mathbb{Z}_{2}$,
likely related to the non-locality of the fractional excitations
\cite{Savary2017}.

Frustration of spin interactions is a prime ingredient to drive magnetic systems
into QSL states. This renders quantum Monte-Carlo (QMC) calculations widely
inapplicable as an unbiased tool, because of the minus-sign problem
\cite{Henelius2000}. Among the few exceptions, in which a $\mathbb{Z}_{2}$ QSLs
can exactly be shown to exists in $D\geq 2$ dimensions \cite{Kitaev2006} and for
which quantum Monte-Carlo methods can be applied \cite{Nasu2014}, is Kitaev's
spin model, with frustrated compass exchange on the honeycomb lattice. Another
case of great interest are variants of the class of Balents-Fisher-Girvin (BFG)
type models \cite{Balents2002}, which comprise certain XXZ Hamiltonians with
Ising frustration and easy-axis anisotropy, e.g. on the kagome lattice. In the
strong-anisotropy limit, the low-energy manifolds of these models are
exponentially degenerate from $S^z$ constraints on local units and feature
ring exchange $K\sim J_{XY}^2/J_z$. The latter can lead to dynamics dual to
quantum dimer models \cite{Rokhsar1988, Moessner2001} and realizes a similar
$\mathbb{Z}_{2}$ gauge structure \cite{Moessner2001a, Sheng2005, Isakov2006a,
Isakov2007, Dang2011, Isakov2011, Isakov2012}.

Consequently the BFG QSLs are topologically ordered with four-fold ground state
degeneracy on the torus in 2D and represent a deconfined phase \cite{Wegner1971,
Wen1991, Kitaev2003}. The deconfined non-local elementary excitations are
analogous to those of the toric code, i.e. 'electric' $e$-particles (spinons,
neutral spin-$1/2$ excitations) and 'magnetic' $m$-particles (vortices, visons)
\cite{Kitaev2003, Savary2017, Moessner2001, Sachdev2018}. The $e$- and
$m$-particles are {\it relative} semions \cite{Kitaev2003, Savary2017}, and in
the BFG models the spinons are known to have bosonic statistics
\cite{Balents2002}. As for additional fingerprints of a $\mathbb{Z}_{2}$ QSL,
the BFG phase hosts symmetry-protected edge states for open boundary conditions
\cite{Wang2017} and displays topological entanglement entropy \cite{Isakov2011,
Isakov2012}.

An explicit low-energy Hamiltonian to study the BFG QSL on the kagome
lattice is the XY-model with ring exchange on four-site bow-ties
\cite{Dang2011, Becker2018}, as in Fig.~\ref{fig:model}
\begin{equation}
H = - \frac{1}{2}\sum_{\Braket{ij}} J_{ij}
\left( S_i^+ S_j^- + h.c. \right)
- K \sum_{\Braket{ijkl} \in \bowtie} P_{ijkl}\,,\label{eq:genh}
\end{equation}
where $S^{\pm}_i$ are spin-1/2 raising and lowering operators on sites $i$ and
$J_{ij}\geq 0$ and $K \geq 0$ are the nearest-neighbor- (NN) and ring exchange
constants, respectively. The ring exchange $P_{ijkl} = S_i^+ S_j^- S_k^+ S_l^- +
S_i^- S_j^+ S_k^- S_l^+ $ acts on each bow-tie ($\bowtie$) as illustrated in
Fig.~\ref{fig:model}(a). This model is amenable to QMC analysis, as it
lacks a minus-sign problem. Variants of it, introducing additional interactions,
including also the $S^z$ components have been considered, focusing either on the
spin, or its equivalent hard-core boson formulation \cite{Sheng2005,
Isakov2006a, Isakov2007, Isakov2011, Isakov2012, Becker2019}.

With nearest-neighbor exchange $J_{ij}=J$, Hamiltonian~(\ref{eq:genh}) has
been shown to harbor two quantum phases versus $J/K$ \cite{Dang2011}. For
$K\rightarrow 0$ the model represents the spin-$1/2$ XY ferromagnet (FM),
displaying a superfluid (SFL) phase in terms of the hard-core boson
language. For $J\rightarrow 0$ a BFG QSL is established. The critical
coupling is $(K/J)_c\approx 10.9$ \cite{Dang2011, NoteDang2011}.
Approaching the critical point out of the QSL, the spinons condense into the
superfluid density $\langle b \rangle$ of the hard-core bosons, undergoing a
conventional XY transition. The latter features standard values for the
exponent $\nu\simeq 0.6717$ of the divergence of the correlation length and
$z=1$ for the dynamical critical exponent, consistent with 3D XY
universality.

Remarkably however, due to the composite nature of the hard-core bosons in
terms of the spinons, the exponent of the equal-time boson-correlation
function turns into a fingerprint of the deconfined quantum critical nature
of the transition \cite{Chubukov1994a}.  In fact, $\eta^\star\approx
1.339-1.493$ has been established \cite{Chubukov1994a, Chubukov1994b,
Senthil2002, Isakov2005, Ballesteros1996, Calabrese2002, Isakov2012}, which
is different from standard 3D XY universality, i.e. $\eta\approx 0.038$
\cite{Campostrini2001}, and therefore is referred to as XY$^\star$.

A natural extension of (\ref{eq:genh}) is to include inhomogeneity of the NN
exchange in terms of {\it trimerization}, also known as {\it breathing}, in
which the spins belonging to upward(downward) facing triangles experience
different exchange couplings, $J_{\vartriangle(\triangledown)}$. Such
generalization is motivated both by theory and experiment. With respect to the
former, recent analysis of other breathing kagome spin systems,
i.e. antiferromagnetic XYZ models indicate that trimerization may help to
stabilize quantum disordered phases and QSLs \cite{Mila1998, Mambrini2000,
Zhitomirsky2005, Schaffer2017, Pollmann2017, Iqbal2018, Iqbal2019}. However
for BFG-type models such analysis is lacking. With respect to experiment,
several interacting hard-core boson systems on breathing kagome lattices
have been realized in recent ultracold-atom systems \cite{Santos2004,
Damski2005, Windpassinger2013, Jo2012, Barter2020}.

In this context, the main purpose of this work is to uncover the quantum
magnetism of the XY model with ring exchange on the breathing kagome lattice
versus the trimerization ratio. Our prime focus will be on the spin stiffness
and the static spin structure factor (SSSF) in order to analyze the shift of the
quantum critical point (QCP) versus $J_\vartriangle / J_\triangledown$. In
addition we will consider the dynamical spin structure factor (DSSF) in order to
shed light on the elementary excitations.

The outline of the paper is as follows. In Section~\ref{sec:model}, we list
several details of the model. Section~\ref{sec:method} sketches the QMC
method. In Section \ref{sec:observe}, the extraction from QMC of observables
relevant to our study is described. Section~\ref{sec:results} is devoted to our
results, comprising the spin stiffness in Section \ref{chap:stiffness}, the
quantum phase diagram in Section \ref{ssec:QPD}, the static and dynamic
structure factors in Sections \ref{ssec:SSSF} and \ref{ssec:DSSF}. We conclude
in Section~\ref{sec:conclusion}. We provide Appendix \ref{sec:rt} on the single
trimer stiffness and Appendix \ref{chap:lswt} on the linear spin wave theory
(LSWT) for the breathing XY kagome ferromagnet.


\section{Model and Method}
\subsection{Model}\label{sec:model}

\begin{figure}[tb]
 \centering
  \includegraphics[width=0.45\textwidth]{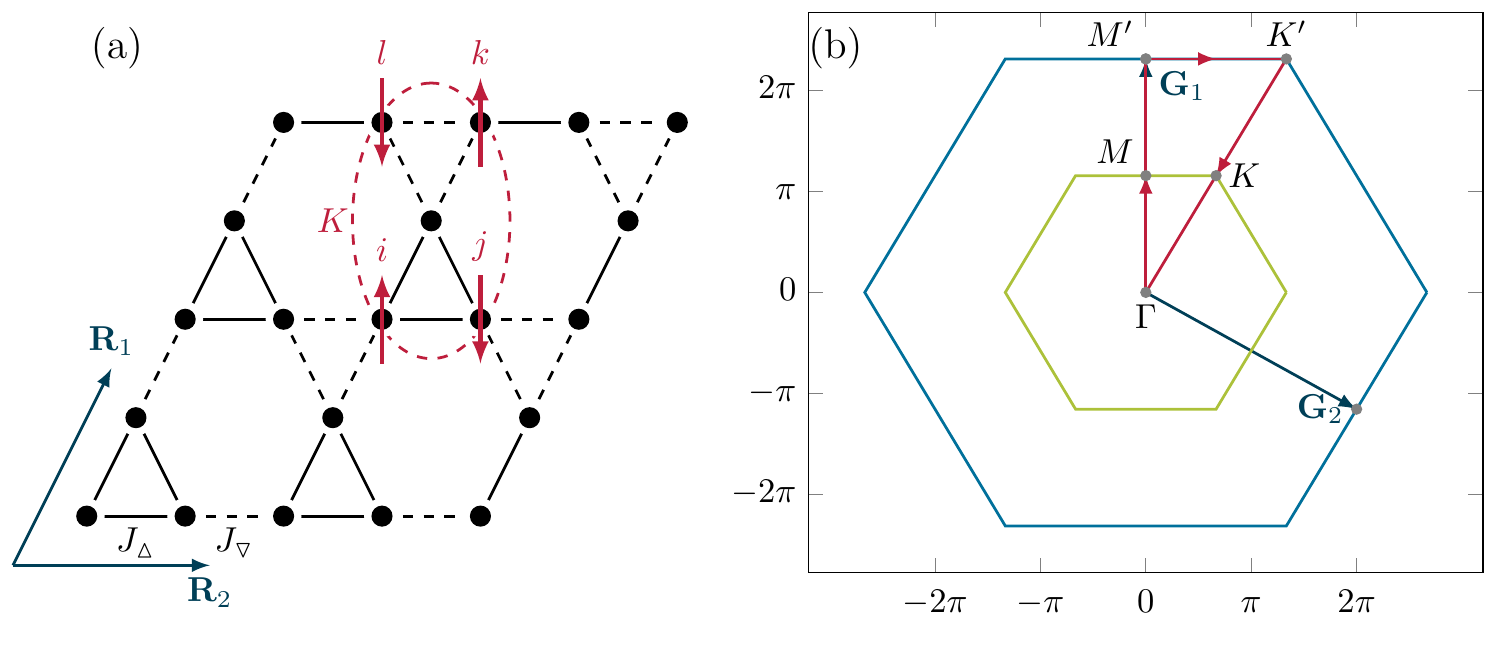}
  \caption{
(a) Trimerized kagome lattice, with NN and ring exchange
$J_{\vartriangle(\triangledown)}$ and $K$. (b) First and second Brillouin zone
 with high-symmetry path.
}
  \label{fig:model}
\end{figure}

The breathing version of model (\ref{eq:genh}) reads
\begin{align}
H = & - \frac{1}{2} ( \sum_{\vartriangle\Braket{ij}} J_\vartriangle S_i^+ S_j^- +
\sum_{\triangledown\Braket{ij}} J_\triangledown S_i^+ S_j^- + h.c. )
\nonumber \\
& - K \sum_{\Braket{ijkl} \in \bowtie} P_{ijkl}\,,
\label{eq:hamiltonian}
\end{align}
where $J_{\vartriangle(\triangledown)}$ refer to the solid(dashed) up(down)
triangles on the kagome lattice, depicted in Fig.~\ref{fig:model}(a),
which is a triangular Bravais lattice with lattice vectors ${\textbf{R}_{1,2} =
( \frac12 , \frac{\sqrt{3}}{2} ), ( 1 , 0 )}$ for a lattice constant
$l{=}1$ hereafter, and a three-site basis at ${\textbf{r}_{\alpha=0,1,2} = (
0 , 0 ), ( \frac14, \frac{\sqrt{3}}{4} ), ( \frac12, 0 )}$, i.e. the intersite
distance is $a=1/2$ and the number of spins is $N=3 L^2$. The
reciprocal lattice vectors are $\textbf{G}_{1,2} = ( 0, \frac{4\pi}{\sqrt{3}} ) ,
( 2\pi, -\frac{2\pi}{\sqrt{3}} )$, with $\textbf{G}_i \cdot \textbf{R}_j = 2\pi
\delta_{ij}$ and the Brillouin zone (BZ) is set by $\textbf{q}= \frac{n_1}{L}
\textbf{G}_1 + \frac{n_2}{L} \textbf{G}_2$ with ${n_{1,2} = 0, 1, \dots, L-1}$,
see Fig.~\ref{fig:model}(b).

For the remainder of the text we will employ a number of dimensionless
parameters, i.e. $j=J_{\triangledown} / J_{\vartriangle}$, $k= K /
J_{\vartriangle}$ and ${\kappa = K / \bar{J}}$. There we define a mean
exchange coupling ${\bar{J} = (J_{\vartriangle} + J_{\triangledown})/2}$ with a
dimensionless version of $\bar{\jmath} = \bar{J}/J_{\vartriangle} =
(1+j)/2$. In this work, we will consider the region of $0 \leq j \leq 1$ and
$k \geq 0$.


\subsection{Quantum Monte-Carlo method}\label{sec:method}

The numerical results of this work are obtained from QMC calculations,
using the stochastic series expansion (SSE) \cite{Sandvik1992, Sandvik1999,
Syljuasen2002}. This method is based on an importance sampling of the high
temperature series expansion of the partition function
\begin{equation}\label{eq:partition}
\mathcal{Z} = \sum_{\alpha}\sum_{S_{M}}\frac{(-\beta)^{n}(M-n)!}{M!}
\langle \alpha|\prod_{a=1}^{M} H_{t_{a},p_{a}}|\alpha\rangle \, ,
\end{equation}
where $\beta{=}1/T$ is the inverse temperature and $M$ the truncation order,
self-adjusted to the desired precision.
As compared to conventional implementations, we enhance the approach by
including the ring exchange following Ref.~\cite{Melko2005}. I.e. the entries
$H_{t_{a},p_{a}}$ of the operator string $\prod_{a=1}^{M}H_{t_{a},p_{a}}$ now
comprise $C$-, $J$-, and $K$-type operators at plaquette~$p$,
${H_{C,p} = C I_{ijkl}}$,
${H_{J,p} = (S_{i}^{+} S_{j}^{-} + S_{i}^{-} S_{j}^{+})I_{kl}/2}$, and
${H_{K,p} = S_{i}^{+} S_{j}^{-} S_{k}^{+} S_{l}^{-} +
S_{i}^{-} S_{j}^{+} S_{k}^{-} S_{l}^{+}}$,
including appropriate permutations of $ijkl$.  $C$ has to be chosen such that
all weights of the $C$-type operators are nonnegative.  ${\Ket{\alpha} =
\Ket{S_{1}^{z}, \ldots,S_{N}^{z}} }$ refers to the $S^{z}$ basis and
${S_{M}=[t_{1},p_{1}][t_{2},p_{2}]\ldots[t_{M},p_{M}]}$ is an index for the
operator string.

This operator string is sampled, using a Markovian-chain Metropolis scheme,
employing three types of updates, i.e. (i) diagonal updates which change the
number of $C$-type operators $H_{C,p}$ in the operator string, (ii) loop
updates which change the type of operators $H_{C,p}\leftrightarrow
H_{J,p}$ and $H_{J,p}\leftrightarrow H_{K,p}$, and (iii)
multibranch cluster update which change the type of operators
$H_{C,p}\leftrightarrow H_{K,p}$. The latter update refers to the
prime new ingredient for ring exchange models \cite{Melko2005}. For bipartite
lattices the loop update comprises an even number of off-diagonal operators
($J$- and $K$-type). This ensures positivity of the transition
probabilities and waives the minus-sign problem.

\subsection{Observables}\label{sec:observe}

Here we briefly sketch formal details regarding the main physical
observables which we evaluate.

\subsubsection{Spin Stiffness}

To obtain the superfluid density, we calculate the spin stiffness (or helicity
modulus) $\rho_S$
\begin{equation}
\rho_S = \left. \frac{\partial^2 F(\phi)}
{\partial \phi^2}\right|_{\phi=0} \, ,
\label{rhos}
\end{equation}
which indicates the presence of long-range order (LRO). Here $F(\phi)$ is the free
energy versus a twist of the spins in the XY-plane, with an angle increasing
by $\phi=\Phi/L$ per bond for any given bond direction. At $T=0$, the free
energy is replaced by the ground state energy $E(\phi)$. We implement the
estimator for this quantity following Refs. \cite{Cuccoli2003, Sandvik1997,
Pollock1987, Harada1998}, using that the r.h.s. of (\ref{rhos}) can be
expressed in term of squares of operators $S_i^+S_j^-$, transporting
$\uparrow$-spins along the $ij$-bond. I.e.
\begin{equation}
 \rho_S = \frac{1}{d \beta} \sum_\alpha^d \langle W_\alpha^2 \rangle \, ,
\label{rhowind}
\end{equation}
where $d=2$ refers to the dimension and the winding number $W_\alpha$ is
defined by
\begin{equation}
W_\alpha = \frac{1}{L_\alpha} \sum_b N_{b,\alpha} \, ,
\label{windn}
\end{equation}
where $\alpha=x,y$ is the spatial direction and $L_\alpha$ the number of bonds
per spatial direction.
The sign of the phase factor is $N_{b,\alpha} = \pm 1$ whether the bond operator
transports a net spin current in $+$ or $-$ spatial direction,
which corresponds to the operators $S^+S^-$ or $S^-S^+$ in the operator string.

\subsubsection{Spin Structure Factors}\label{chap:SSF}

We evaluate two types of spin correlation functions. First, the longitudinal,
i.e $zz$, static spin structure factor (SSSF) \cite{Dang2011}
\begin{equation}
S({\bf q})  = \frac{1}{N}
\sum_{i\alpha,j\beta} e^{\mathrm{i} {\bf q} \cdot ({\bf R}_{i\alpha} -
{\bf R}_{j\beta}) }
\langle S_{i\alpha}^{z}S_{{j\beta}}^{z} \rangle \, ,
\label{sssf}
\end{equation}
where $\langle S_{i\alpha}^{z}S_{{j\beta}}^{z} \rangle$ is extracted during
the simulation, and the sites $i\alpha$ correspond to lattice coordinates
${{\bf R}_{i\alpha}{=} n_{1i} {\bf R}_1 {+} n_{2i} {\bf R}_2 {+} {\bf r}_\alpha}$,
with $n_{1i(2i)} {=} 0, 1, \dots, L-1$.

Additionally we evaluate the transverse, i.e. $+{-}$, dynamic spin structure
factor (DSSF). In real space and at imaginary time $\tau$ this is obtained from
the SSE by \cite{Sandvik1992}
\begin{align}
\Braket{S^+_i(\tau)S^-_j(0)} =
& \left\langle \sum_{m=0}^{M} {M \choose m}
\left(\frac{\tau}{\beta}\right)^m\nonumber
\left(1-\frac{\tau}{\beta}\right)^{M-m}\nonumber \right. \\
& \quad \left. \frac{1}{M} \sum_{p=0}^{M-1} S_{i}^{+}(m+p)S_{j}^{-}(p)
\right\rangle_{W} \, ,
\label{a2}
\end{align}
where $i,j$ refer to any lattice site,  $m+p, p$ label positions within the
operator string, i.e. intermediate time slices, and $\langle \dots
\rangle_{W}$ denotes the Metropolis weight of an operator string of length $M$
generated by the SSE \cite{Sandvik1999, Syljuasen2002}.

From \eqref{a2} we proceed to momentum space by Fourier transformation and define
\begin{equation}\label{eqn:structureFactor}
S_{\alpha\beta}({\bf q}, \tau) = \frac{1}{N}
\sum_{i,j }e^{\mathrm{i} {\bf q} \cdot ({\bf R}_{i\alpha} - {\bf R}_{j\beta})}
\Braket{S^+_{i\alpha}(\tau) S^-_{j\beta}(0)} \, ,
\end{equation}
which due to the three-sites basis is a $3 \times 3$ matrix. To
analyze the excitation spectrum, we follow Ref. \cite{Becker2018} and confine
ourselves to the analytic continuation of the trace $\sum_\alpha
S_{\alpha,\alpha}({\bf q}, \tau)$. This trace is invariant with respect to
any unitary transformation of $S_{\alpha,\beta}({\bf q}, \tau)$ and in
particular we may rotate onto its principal axis. I.e. we diagonalize this
matrix for each ${\bf q}$ and $\tau$ and transform to real frequencies
$\omega$ for each of the eigenvalue $S_\mu({\bf q}, \tau)$, $\mu$=0,1,2
separately
\begin{equation}
S_\mu({\bf q}, \tau) =\int_{0}^{\infty}
d\omega \, S_\mu({\bf q} ,\omega)K(\omega,\tau) \, ,
\label{ac}
\end{equation}
with a kernel $K(\omega,\tau)=(e^{-\tau\omega}+e^{-(\beta-\tau)\omega})/\pi$.
We find, the eigenvalue decomposition to allow for a higher
resolution in the analytic continuation, if regions of dominant spectral
features occur nearby in energy space. In principle, only $\sum_\mu
S_\mu({\bf q} ,\omega)$ refers to the actual total spectrum and the $S_\mu({\bf q}
,\omega)$ are auxiliary functions, which do {\em not} necessarily encode any
decomposition into sharp eigenmodes. Yet, visualizing
each $S_\mu({\bf q} ,\omega)$ can be instructive.

The inversion (\ref{ac}) is an ill-posed problem, for which maximum entropy
methods (MEM) have proven to be well suited. Here we use Bryan's MEM
algorithm \cite{Skilling1984, Bryan1990, Jarrell1996}. This method minimizes
the functional $Q=\chi^{2}/2-\alpha\sigma$, with $\chi$ being the covariance
of the QMC data with respect to the MEM trial spectrum $S_\mu({\bf q}, \omega)$.
Overfitting is prevented by an entropy term $\sigma=\sum_{\omega} S_\mu({\bf q},
\omega) \ln[S_\mu({\bf q}, \omega) /m(\omega)]$.  We have used a flat default
model $m(\omega)$, which is iteratively adjusted to match the zeroth moment of
the trial spectrum.  The optimal spectrum follows from the average of $S_\mu({\bf
q}, \omega)$, weighted by a probability distribution $P[\alpha| S_\mu({\bf q},
\omega)]$ \cite{Skilling1984}.


\section{Results}\label{sec:results}

In this section, we detail our findings for the spin stiffness and the quantum
phase diagram, as well as the static, and dynamic structure factors in order
to characterize the phases of the model versus ring exchange and
trimerization.

\subsection{Spin Stiffness}
\label{chap:stiffness}

\begin{figure}[tb]
 \centering
  \includegraphics[width=0.45\textwidth]{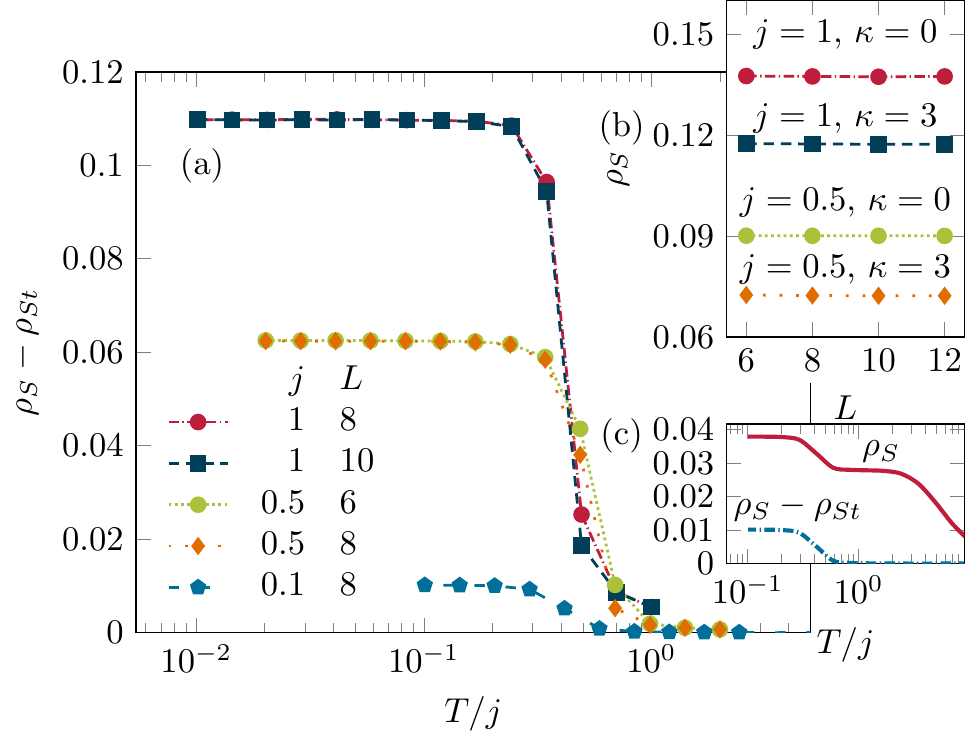}
  \caption{
(a) Spin stiffness $\rho_S$ for $\kappa=0$, subtracted by single trimer
spin stiffness $\rho_{St}$, for various trimerization ratios $j=1,0.5,0.1$ shown
over wide temperature range.
Low-temperature plateaus set the scale $T^\ast = j/L$.
Note: $x$-axis scaled by $j$.
(b) Inset shows spin stiffness versus $L=6,\dots,12$ at fixed low temperature
$T=j/10$, for various trimerization ratios $j=1,0.5$ and ring exchange $\kappa=0,3$.
Errorbars smaller than marker size.
(c) Inset shows spin stiffness $\rho_S$ and $\rho_S -\rho_{St}$ at $L=8$
and $j=0.1$ versus temperature.
}
  \label{fig:spinStiffnessT}
\end{figure}

We use the stiffness to extract the quantum critical point. To this end we
rely on the fact that scaling theory \cite{Fisher1989} has been proven to
apply for $j=1$ \cite{Dang2011} and we anticipate that chosing $j\neq 1$
should not change this. This implies that the stiffness fulfills the scaling
relation $\rho_S = L^{-z} F ( (\kappa_C-\kappa) L^{1/\nu}, \beta/L^z )$ with
a universal scaling function $F$. In turn, using $z=1$ for the dynamical
critical exponent, and fixing the temperature such that $L^z/\beta = c$ is
constant, i.e. $T = T^\star = c/L$, the function $\rho_S L$ will collapse
onto a single curve versus $(\kappa_C-\kappa) L^{1/\nu}$ for all $L$.
We use a small constant $c$, such that $\rho_S$ represent the zero
temperature limit for $T<T^\star$ for each $L$ studied. Furthermore, since
$\rho_S(\kappa=\mathrm{const.},L)=\mathrm{const.}$ for $L\gg 1$ and
$T\rightarrow 0$, this implies that $F(x,\mathrm{const})\propto x^\nu$.

In Fig. \ref{fig:spinStiffnessT}(a), and prior to analyzing the
scaling behavior, we display the spin stiffness versus temperature for
various trimerization at $\kappa=0$ where LRO is certain at $T=0$ for
$j=1$. For $j\rightarrow 0$, a stiffness $\rho_{St}$ of isolated trimers
remains, which we subtract off for simplicity. The impact of retaining
$\rho_{St}$ is exemplified for $j=0.1$ in Fig. \ref{fig:spinStiffnessT}(c).
Several points can be read off from this figure. First, there is a clear
low-$T$ crossover regime to a state with a finite stiffness, which decreases
with increasing trimerization. Choosing $T^\star$ well below this regime is
sufficient for the scaling analysis to describe zero temperature
behavior. Second, the crossover regimes for different $j$ can be made to
coincide, if $T$ is scaled by $j$. Third, the figure shows that the crossover
regime is rather insensitive with respect to $L$ for the systems sizes we
have investigated. Therefore, we use $c\simeq j$, i.e. $T^\star =
j/L$ as a reasonable choice to obtain quasi ground state properties.

We also note from Fig.~\ref{fig:spinStiffnessT}(a), that the low-$T$ saturation
value of $\rho_{S}-\rho_{St}$ depends little on system size. Sufficiently close
to criticality, this is to be expected from scaling. I.e., for any
temperature on the low-$T$ saturation plateau of $\rho_{S(t)}$, the universal
function $F(x,y)$ is in its asymptotic regime $F(x,\mathrm{const})\propto
x^\nu$. At fixed $\kappa_{(C)}$ therefore, $\rho_{S}\propto
L^{-z}(L^{1/\nu})^\nu = \mathrm{const}$ versus $L$. For $\rho_{St}$ the latter
is satisfied trivially. The inset Fig. \ref{fig:spinStiffnessT}(b) demonstrates,
that the independence of system size of $\rho_{S}$ at low $T$ remains valid over
a wide range of $j$- and $\kappa$-values, irrespective of criticality, rendering
the usage of the scaling function rather robust.

\begin{figure}[tb]
 \centering
  \includegraphics[width=0.45\textwidth]{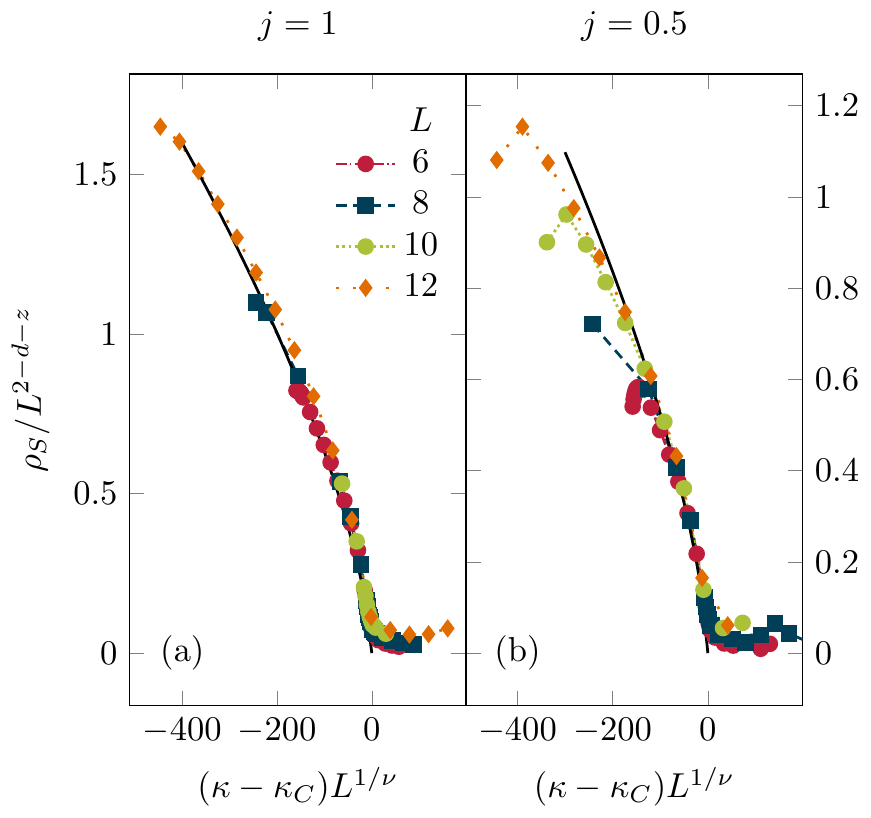}
  \caption{
Scaling behavior at $T^\ast=j/L$ of $\rho_S/L^{2-d-z}$ versus
$(\kappa-\kappa_C)L^{1/\nu}$ at (a) $j=1$ and (b) $j=0.5$.
For $j=1$ ($j=0.5$) QMC data collapses at
$\kappa_C \approx 11.04$ ($\kappa_C \approx 10.97$).
Solid black curve $\propto (\kappa_C - \kappa)^\nu$.
}
  \label{fig:scalingBehavior}
\end{figure}

Setting $T=T^\star$, we now extract the QCP by optimizing the collapse of all
of our results for $\rho_S L^z$ versus $(\kappa_C-\kappa) L^{1/\nu}$ for $L=6
\dots 12$ at fixed $j$, for various $j=0.1 \dots 1$ and employing the 3D
XY$^\star$ universality established, i.e. $z=1$ and $\nu=0.6717$. This
leaves $\kappa_C$ as the single fit parameter.

Results for this optimization procedures are depicted in
Fig.~\ref{fig:scalingBehavior} for two different $j$. Similar behavior is
obtained for all other $j$ considered. The collapse is clearly evident and
leads to a critical ring exchange $\kappa_C\approx 11.04$ and $10.97$ at
$j=1$ and $0.5$, respectively. The value for $\kappa_C(j=1)$ agrees with
previous findings \cite{Dang2011}. The figure displays an additional function
$\propto (\kappa-\kappa_C)^\nu$, which can be superimposed to fit the QMC
results very well in the ordered phase. Apart from the collapse itself, this
provides further support for the validity of the scaling hypothesis also at
finite trimerization. In the LRO phase at $j\neq 1$, and significantly off
from criticality, non-universal corrections in terms of an intermediate
maximum in $\rho_S L$ appear at finite $\kappa$. We observe this maximum at
all $j\neq 1$ investigated. Yet, since $\rho_S$ remains finite, the
stiffness provides no evidence for phases other than the SFL and QSL at $j\neq
1$.

\begin{figure}[tb]
\begin{centering}
\includegraphics[width=0.75\columnwidth]{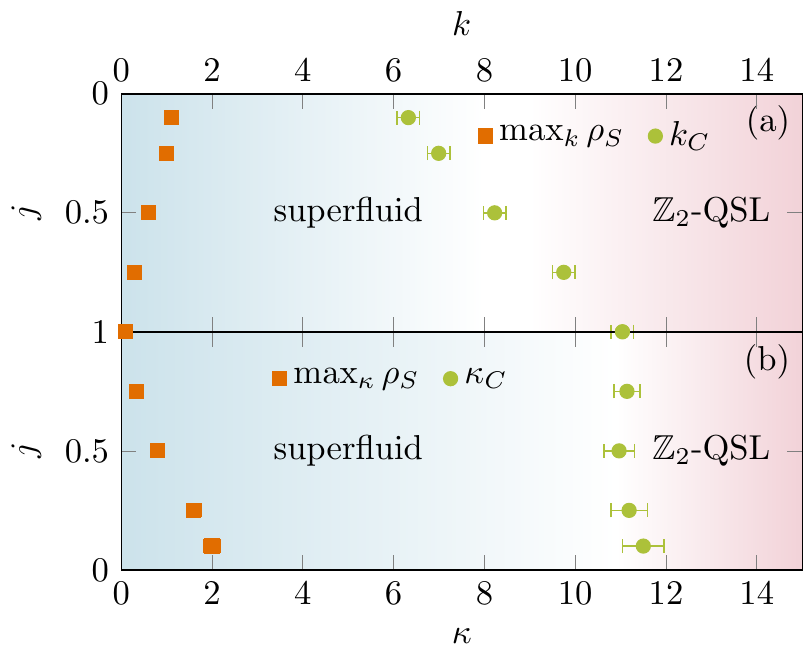}
\end{centering}
\caption{\label{fig:qpd}
Quantum phase diagram versus (a)  ring exchange $k$,
(b) dimensionless ring exchange $\kappa$ and trimerization ratio $j$.
Green symbols, critical coupling (a) $k_C$, (b) $\kappa_C$ for SFL-QSL QCP.
Orange squares, location of $\rho_S$-maximum.
If not visible, errrobars smaller than marker size.
Note: opposite $y$-axis directions in (a) and (b).
}
\end{figure}

\subsection{Quantum Phase Diagram}\label{ssec:QPD}

Using all $\kappa_C(j)$ obtained by the data collapse, we are in a position to
extend the quantum phase diagram (QPD) derived in Ref.~\cite{Dang2011} from the
line $j=1$ onto the $(\kappa,j)$-plane. This is shown in
Fig.~\ref{fig:qpd}(b). Remarkably, if expressed in terms of $\kappa = 2 K /
(J_{\vartriangle} + J_{\triangledown}) = 2 (K/J_{\vartriangle}) / (1+j)$ the
transition resides at a fixed location versus $j$ within the error of the data
collapse. Speaking differently, and equally remarkable, removing the implicit
$j$-dependence from the $x$-axis, as in Fig.~\ref{fig:qpd}(a),
the QPD reveals an increase in the tendency to form the QSL as the trimerization
increases. This can be understood by the
decrease with $j$ of the boson kinetic energy which drives the SFL phase.
In fact, performing linear spin-wave theory (LSWT) exactly at $K=0$, see
Appendix~\ref{chap:lswt}, each spin is connected to four neighboring ones, two
of them by $J_{\vartriangle}$ and two of them by $J_\triangledown$. Therefore
the leading order-$1/S$ contribution to the energy is proportional to
$\bar{\jmath}$, see Eqns.~(\ref{eq:c5}, \ref{eq:c6}).

It is tempting to speculate about the QPD as $j\rightarrow 0$. First,
convergence issues with the QMC prevent us from studying $j=0$ directly. Second,
strictly at $j=\kappa=0$, there can be no SFL LRO phase, since the system is a
lattice of disconnected trimers. Finally, it is likely, that the QSL remains
existent for $\kappa>\kappa_C$, also at $j=0$. For $\kappa<\kappa_C$, one
scenario could be that the superfluid density vanishes as $j\rightarrow 0$,
consistent with Fig.~\ref{fig:spinStiffnessT}, such that ${\rho_S \leq
\rho_{St}}$ on the line $j=0$ for all $\kappa<\kappa_C$. However, this renders
the nature of the state for $\kappa<\kappa_C$ at $j=0$ unclear. Another
scenario, suggested by the maximum in $\rho_S$ could be, that weak LRO, driven
by ring exchange, remains present even at $j=0$, but with a non-monotonous
behavior versus $\kappa<\kappa_C$. This remains an open issue.

\subsection{Static Spin Structure Factor}\label{ssec:SSSF}

\begin{figure}[tb]
\centering
\includegraphics[width=0.9\columnwidth]{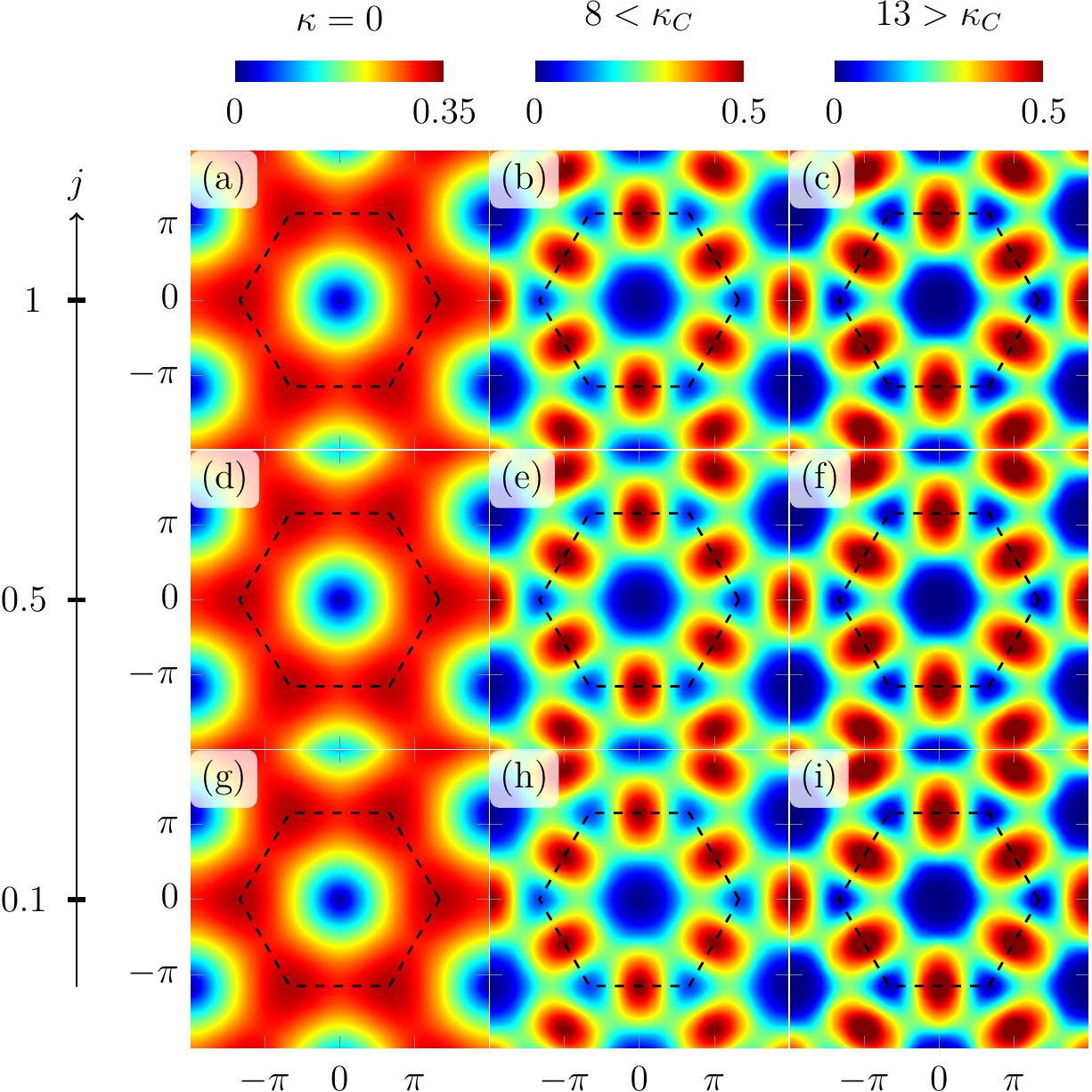}
\caption{SSSF $S({\bf q})$ for various trimerization ratios
${j=1,0.5,0.1}$ and ring exchange couplings $\kappa=0$, $8(<\kappa_C)$, and
$13(>\kappa_C)$. QMC at $L=12$, $T=T^{\ast}$.
}
\label{fig:StaticSF}
\end{figure}

The SSSF $S({\bf q})$ is equivalent to the static density-density correlation
function within the hard-core boson picture.
As such it has been used to check if the transition from the SFL into the
anticipated $\mathbb{Z}_2$-QSL is accompanied
by a density wave, i.e. by any peaks scaling $\propto N$. This has been excluded
in \cite{Dang2011} at $j=1$, proving that the non-SFL phase does not break
discrete lattice symmetries and thereby further corroborating its QSL nature.

In view of the QPD in Fig.~\ref{fig:qpd}, the short-range nature of the
$zz$-spin-correlations will remain true also for $j<1$. Therefore, rather than
repeating the finite-size scaling analysis of \cite{Dang2011}, we highlight
instead, that the SSSF is almost insensitive to trimerization if considered in
terms of the rescaled ring exchange $\kappa$. This is shown in
Fig.~\ref{fig:StaticSF}, which depicts contours of $S({\bf q})$ at $T=T^\ast$
and for $L=12$ at $\kappa=0$, $8(<\kappa_C)$, and $13(>\kappa_C)$.

Indeed, the changes in this figure along the vertical $j$-direction in each
column are small only. The evolution of the contours along the
$\kappa$-direction within the $j=1$ row are consistent with that in
\cite{Dang2011}. For $\kappa\ll 1$, the maximum intensity in $S({\bf q})$
resides at the ${\bf K}$-points, due to the NN-correlations from $j$.
As $\kappa$ increases, the next NN-correlations produce maximum
intensity at the ${\bf M}$-points and minimum intensity at the ${\bf \Gamma}$-
and ${\bf K}$-points.

\begin{figure}[tb]
\centering
\includegraphics[width=0.45\textwidth]{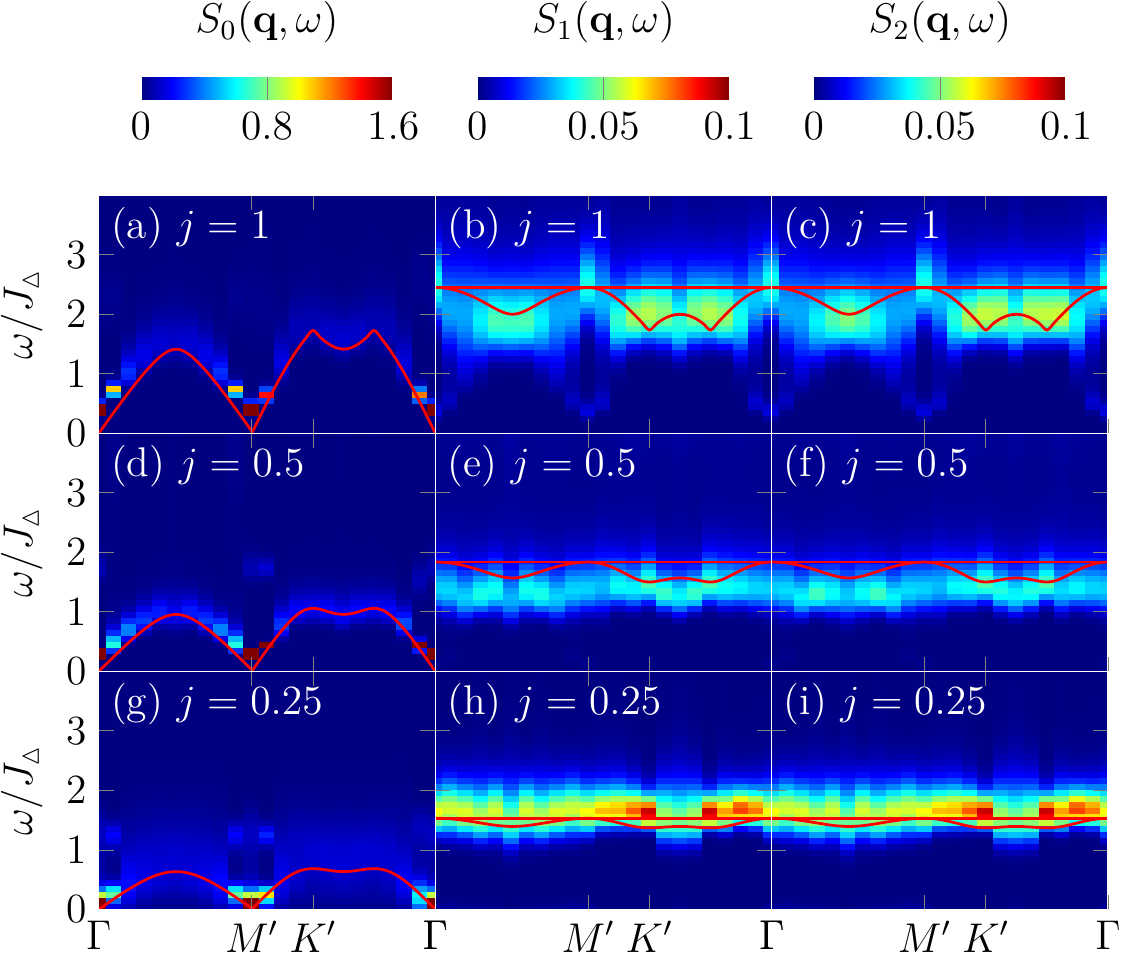}
\caption{
DSSF in SFL phase at $\kappa=0$ for three trimerization ratios ${j=1,0.5,0.25}$
(rows), along high-symmetry path Fig.~\ref{fig:model}(b) in first and second BZ.
Columns refer to the three diagonal modes $S_\mu({\bf q} ,\omega)$ in
Eq.~\eqref{ac}. $\sum_\mu S_\mu({\bf q},\omega)$ yields total spectrum.
QMC at $L=12$, $T=T^\ast$.  Solid red line: LSWT dispersions
Eqns.~(\ref{eq:c5}, \ref{eq:c6}).
}
\label{fig:SpectrumFM}
\end{figure}

\subsection{Dynamic Spin Structure Factor}\label{ssec:DSSF}

As to be expected from the QPD, the elementary excitations fall into one of two
classes. I.e. for $\kappa{<}\kappa_C$ they are magnons of a planar FM, with a
broken residual $U(1)$-symmetry and for $\kappa{>}\kappa_C$ they represent the
deconfined spinons and visons of the $\mathbb{Z}_2$-QSL. While the DSSF can be
used directly to map out the magnon spectrum, it cannot do so for the
fractionalized one-particle excitations. However, since the DSSF comprises a
two-particle correlation function in terms of the latter, it can access the
two-spinon spectrum, which is a continuum at each fixed total momentum. Using
QMC at $j=1$, magnons (two-spinon continua) have been verified in the FM
(QSL) phase directly for Hamiltonian (\ref{eq:hamiltonian}) \cite{Becker2019},
and for a closely related model of the BFG class \cite{Becker2018}.
Here we consider the DSSF for $j\leq 1$. Primarily, we focus on the magnon
excitations at $\kappa=0$, since the dynamics in the QSL phase is driven by the
ring exchange, where effects of the trimerization are not expected to be
significant.

\begin{figure}[tb]
\centering
\includegraphics[width=0.45\textwidth]{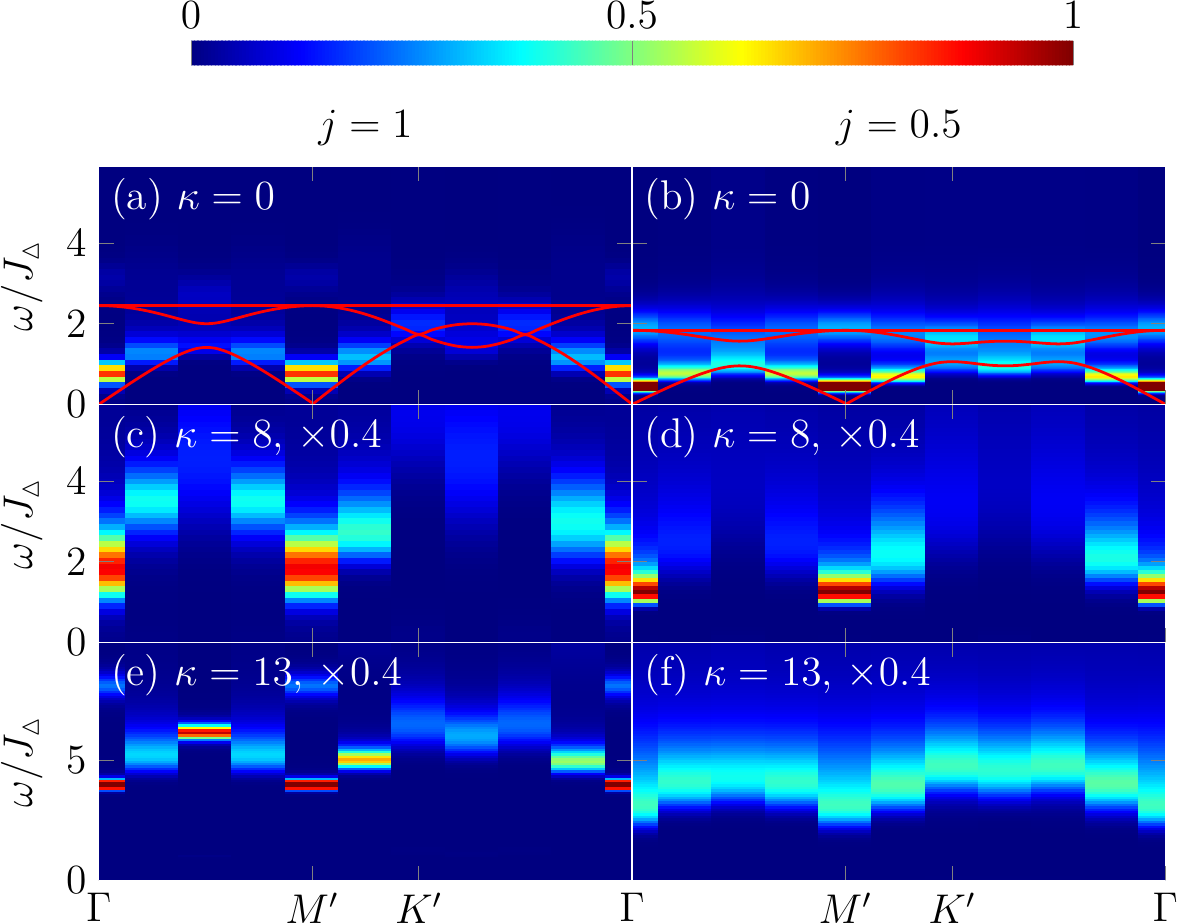}
\caption{
DSSF total spectrum $\sum_\mu S_\mu({\bf q},\omega)$ for three ring exchange
values, in SFL phase at $\kappa=0,8$ (first, second row), and QSL phase at
$\kappa=13$ (third row) for two trimerization ratios $j=1, 0.5$, along high-symmetry
path Fig.~\ref{fig:model}(b) in first and second BZ. QMC at $L=6$ and $T=T^\ast$.
Intensities rescaled for visibility.  Solid red line in first row: LSWT dispersions
Eqns.~(\ref{eq:c5}, \ref{eq:c6}).
}
\label{fig:SpectrumRing}
\end{figure}

Fig.~\ref{fig:SpectrumFM} details the evolution with $j$ of the spectra of
the three auxiliary diagonal modes $S_\mu({\bf q} ,\omega)$ of the
DSSF at $\kappa=0$, with the total spectrum being $\sum_\mu S_\mu({\bf q}
,\omega)$. The dispersions are shown along the high symmetry path of the
extended BZ depicted in Fig.~\ref{fig:model}(b). Each rows of the figure
display the typical signatures of a FM with a three-site unit
cell. I.e. there is one "acoustic" Goldstone magnon and two "optical" magnon
branches. Properly, the dominant spectral weight occurs in the Goldstone mode
at the ordering vector, i.e. the ${\bf \Gamma}$-point. The effect of the
trimerization can be seen clearly along the columns of
Fig.~\ref{fig:SpectrumFM}, i.e. as $j$ is decreased, implying a reduction of
the magnon kinectic energy, the bandwidth for both types of magnons shrinks
accordingly, with however the Goldstone nature of the acoustic mode remaining
intact.

To obtain a qualitative account of the magnon energies to leading $O(1/S)$ we
have performed LSWT. See Appendix~\ref{chap:lswt}. The LSWT dispersions are
included in Fig.~\ref{fig:SpectrumFM}.  Regarding the Goldstone mode, the LSWT
dispersion agrees remarkably well with $S_0({\bf q}, \omega)$ from the QMC.
For the two optical modes of the LSWT, $S_1({\bf q}, \omega)$ and $S_2({\bf
q},\omega)$ certainly are within their energy range, however they do not
separate into distinct branches. Irrespective of the qualitative agreement
between QMC and LSWT, the figure also clearly demonstrates, that at higher
energies the QMC spectra tend to broaden, hinting at the relevance of
magnon-magnon interactions.

Fig.~\ref{fig:SpectrumRing} depicts the evolution of the DSSF with $\kappa$
crossing over from the SFL into the QSL phase, along two lines with $j=1$ and
0.5. Each spectrum refers directly to the trace over the three diagonal
modes. The values of $\kappa$ are chosen to be deep in the SFL phase,
$\kappa=0$, as well as closer to the QCP, both, in the SFL and QSL phase,
i.e. at $\kappa=8$ and 13, respectively. Several points should be
noted. First, in the SFL phase at $\kappa=0$ and 8, the spectra corroborate a
clear Goldstone mode behavior, with a pronounced spectral weight at the ${\bf
\Gamma }$-point. Second, in the QSL phase, the spectral weight may show
remnants of this at $j=1$. However, weight is also redistributed into
other regions of momentum space. For $j=0.5$, the DSSF in the QSL regime is
rather featureless versus momentum, which has to be attributed in part to
less statistics for the imaginary time data in this regime of small $j$ and
large $\kappa$. Third, introducing ring exchange, the spectrum is shifted to
higher energies and is broadened (note the different $y$-scales of the
panels).  Fourth, Fig.~\ref{fig:SpectrumRing} displays an opening of a gap at
the ${\bf \Gamma }$-point as $\kappa$ increases. It is tempting to associate
this with the onset of the two-spinon gap for $\kappa>\kappa_C$. To
substantiate this, larger systems are required to prove a vanishing gap for
$\kappa<\kappa_C$. This is beyond our analysis.


\section{Conclusion}\label{sec:conclusion}

To summarize, using extensive quantum Monte-Carlo calculations, we have
investigated the role of trimerization in the frustrated planar XY spin-$1/2$
ferromagnet on the kagome lattice with four-site ring exchange. Among our main
findings is that qualitatively, several aspects of the physics of this system
already known for the uniform case, are nearly invariant with respect to
trimerization, if the exchange parameters are rescaled properly. This pertains to
the quantum phase diagram, comprising LRO and $\mathbb{Z}_{2}$ QSL states, the
scaling behavior and the universality class, as well as the static structure
factor of the model. Regarding the impact of trimerization on the excitations,
we find that linear spin waves remain a reasonably valid description in the LRO
phase, while spinon continua in the QSL phase may even be intensified in the
non-uniform case. Open questions remain, regarding a non-monotonous behavior of
the spin stiffness with ring exchange at strong trimerization, which may signal
new types of ground states.
Regarding actual realizations in optical lattices or local-moment systems,
our study shows that, in terms of the \textit{boson} exchange parameters,
trimerization enlarges the QSL regime.

\begin{acknowledgments}
Work of N.C. and W.B. has been supported in part by the State of Lower Saxony
through QUANOMET (project NP-2). Work of W.B. has been supported in part by the
DFG through project A02 of SFB 1143 (project-id 247310070). W.B. also
acknowledges kind hospitality of the PSM, Dresden. This research was supported
in part by the National Science Foundation under Grant No. NSF PHY-1748958.
\end{acknowledgments}


\appendix

\section{Spin Stiffness of XY Trimer}\label{sec:rt}

\begin{figure}[b]
\centering
\includegraphics[width=0.45\textwidth]{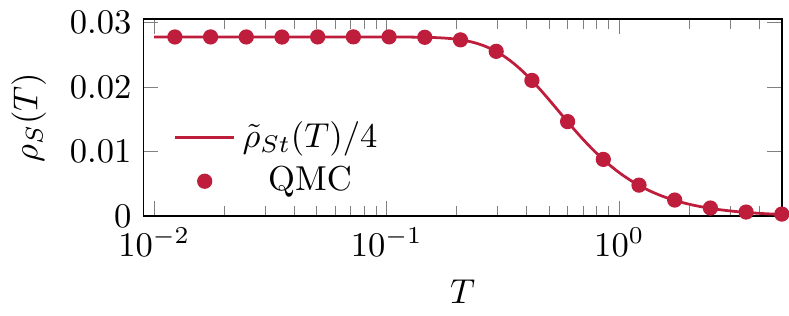}
\caption{
Comparison of spin stiffness versus T for single XY trimer
$\tilde{\rho}_{St}(T)/4$ from Eq.~(\ref{LVA_ThermoQstat}) with QMC trimer
result.
Errorbars smaller than marker size.
}
\label{fig:trimer}
\end{figure}

At $j=k=0$, our model simplifies to disconnected XY trimers on a triangular
lattice. In turn, even in this limit the model exhibits a finite extensive spin
stiffness, resulting from the trimers. Since this does not imply a collective
LRO state, we analyze the stiffness obtained from QMC by subtracting the
bare trimer stiffness. The latter can be obtained analytically by introducing a
twist $\phi$ along the $x$-direction of the trimer. This affects only the $S^z=\pm
1/2$ sectors, containing cyclic permutation of $\Ket{\uparrow \downarrow
\downarrow}$ and $\Ket{\downarrow \uparrow \uparrow }$. For the $S^z=1/2$
\begin{align}
H^{S^z=1/2}_\text{trimer}(\phi) = -\frac{J_{\vartriangle}}{2}
\phantom{aa}
\begin{blockarray}{cccc}
\Ket{\uparrow \uparrow \downarrow} & \Ket{\uparrow \downarrow \uparrow} &
\Ket{\downarrow \uparrow \uparrow} \\
\begin{block}{(ccc)c}
0 & 1 & 1 \\
1 & 0 & e^{\mathrm{i} \phi} \\
1 & e^{-\mathrm{i} \phi} & 0 \\
\end{block}
\phantom{\Ket{\uparrow \uparrow \downarrow}} &
\phantom{\Ket{\uparrow \downarrow \uparrow}} &
\phantom{\Ket{\downarrow \uparrow \uparrow}} \\
\end{blockarray}
\label{eqn:trimerMatrix}
\end{align}
\vspace*{-20pt}\\
and identically for $S^z=-1/2$. Moreover $H_\text{trimer}(\phi)$
$\Ket{\uparrow \uparrow \uparrow (\downarrow \downarrow \downarrow)}$ = $0
\Ket{\uparrow \uparrow \uparrow (\downarrow \downarrow \downarrow)}$. The
stiffness is obtained from the free energy by $\tilde{\rho}_{St}(T)= \partial^2
F(\phi)/\partial \phi^2|_{\phi=0}$. Straightforward algebra yields
\begin{equation}
\tilde{\rho}_{St}(T)=\left[9+\frac{9 (2 (e^{\frac{1}{2 T}}+3)
T+3)}{2 (e^{\frac{3}{2 T}}-1) T-3}\right]^{-1} \, .
\label{LVA_ThermoQstat}
\end{equation}
Fig. \ref{fig:trimer} depicts $\tilde{\rho}_{St}(T)/4$, where the factor of
$4$ refers to $L_\alpha^2$ in Eq.~\eqref{windn} for trimers on a
non-interacting kagome lattice with $j=0$, where $L_\alpha=2$ because of the
$J_{\vartriangle}$ and (vanishing) $J_{\triangledown}$ bonds per spatial
direction.
The figure also proves that QMC data for $\rho_{St}(T)$
for a single trimer from Eqns. (\ref{rhowind}, \ref{windn}) agrees with
Eq. (\ref{LVA_ThermoQstat}).

\section{Linear Spin Wave Theory}\label{chap:lswt}

In the SFL phase and for $\kappa=0$, the QMC spectra can be contrasted against
magnon excitations obtained from linear spin wave theory (LSWT). For the latter
we use the conventional Holstein-Primakoff representation with a quantization
axis along $x$ and expanded up to leading order $1/S$, i.e.
\begin{align}\nonumber
S_m^x & = S - a_m^\dagger a_m \\
S_m^y & \approx \frac{\sqrt{2S}}{2 \mathrm{i}}
\left( a_m - a_m^\dagger \right) \label{eq:HolsteinPrimakoff}
\end{align}
with Boson operators $a_m^{(\dagger)}$. Inserting this into
Eq.~(\ref{eq:hamiltonian}) and after Fourier transformation we get
\begin{align}
& \mathcal{H}^{(2)} = S \sum_{\textbf{q}}
& \Psi_{\textbf{q}}^\dagger
\underbrace{\begin{pmatrix} z \bar{J} \cdot \mathbb{1} - \Gamma(\textbf{q}) &
\Gamma(\textbf{q}) \\
\Gamma(\textbf{q}) & z \bar{J}
\cdot \mathbb{1} - \Gamma(\textbf{q})
\end{pmatrix}}_{M(\textbf{q})} \Psi_{\textbf{q}} \, ,\label{eq:LSWTHamiltonian}
\end{align}
where ${z=z_{\vartriangle} + z_{\triangledown}= 2+2 = 4}$ is the coordination
number, ${\Psi_\textbf{q} = ( a_{0,\textbf{q}}^\dagger \,
a_{1,\textbf{q}}^\dagger \, a_{2,\textbf{q}}^\dagger \, a_{0,\textbf{q}} \,
a_{1,\textbf{q}} \, a_{2,\textbf{q}})}$ is a spinor of creation and annihilation
operators, with $0,1$, and $2$ referring to kagome basis of the triangular
lattice, and $\Gamma(\textbf{q})$ encodes the hopping matrix elements
\begin{widetext}
\begin{align}
\Gamma(\textbf{q}) & = \frac12 \begin{pmatrix}
0 &
J_{\vartriangle} e^{\mathrm{i} \textbf{q} \cdot \textbf{r}_1} +
J_{\triangledown} e^{-\mathrm{i} \textbf{q} \cdot \textbf{r}_1} &
J_{\vartriangle} e^{\mathrm{i} \textbf{q} \cdot \textbf{r}_2} +
J_{\triangledown} e^{-\mathrm{i} \textbf{q} \cdot \textbf{r}_2} \\
J_{\vartriangle} e^{-\mathrm{i} \textbf{q} \cdot \textbf{r}_1} +
J_{\triangledown} e^{\mathrm{i} \textbf{q} \cdot \textbf{r}_1} &
0 &
J_{\vartriangle} e^{\mathrm{i} \textbf{q} \cdot (\textbf{r}_2 -\textbf{r}_1)} +
J_{\triangledown} e^{-\mathrm{i} \textbf{q} \cdot(\textbf{r}_2 -\textbf{r}_1)}\\
J_{\vartriangle} e^{-\mathrm{i} \textbf{q} \cdot \textbf{r}_2} +
J_{\triangledown} e^{\mathrm{i} \textbf{q} \cdot \textbf{r}_2} &
J_{\vartriangle} e^{-\mathrm{i} \textbf{q} \cdot (\textbf{r}_2 -\textbf{r}_1)} +
J_{\triangledown} e^{\mathrm{i} \textbf{q} \cdot(\textbf{r}_2 -\textbf{r}_1)} &
0\\
\end{pmatrix} \, .
\end{align}
\end{widetext}
The magnon dispersions result from the paraunitary secular equation ${\det | S
\Lambda \cdot M(\textbf{q}) - \omega \mathbb{1} | = 0}$, where ${\Lambda =
\left(\begin{smallmatrix} \mathbb{1} & \mathbb{0} \\ \mathbb{0} & - \mathbb{1}
\end{smallmatrix} \right)}$. We find
\begin{align}
\omega_{0,1}^2 / J_{\vartriangle}^2 & = \frac{\bar{\jmath}}{2}
\left[ 6 \bar{\jmath} \mp
\sqrt{(3j-1)^2 + 8 [ 1 + j \, \gamma({\bf q})]} \right]
\label{eq:c5}
\\
\omega_{2}^2 / J_{\vartriangle}^2 & = \frac32 (1+j)^2
= 6 \bar{\jmath}^2
\label{eq:c6}
\end{align}
with $\gamma({\bf q}) = \cos (2{\textbf{q} {\cdot} \textbf{r}_1}) + \cos(2
{\textbf{q} {\cdot} \textbf{r}_2}) + \cos(2 {\textbf{q} {\cdot} (\textbf{r}_2 {-}
\textbf{r}_1)})$.
In Eq. (\ref{eq:c5}), $\omega_0$ corresponds to the minus sign on the right hand
side and the three branches $\omega_0$ and $\omega_{1,2}$ comprise one acoustic
and two optical modes. One of the latter is completely flat, i.e. at $\omega_2$
the magnons are localized from local interference effects, which are a
reoccurring theme for NN-hopping models on the kagome lattice. At the ${\bf
K}$-point, the gap between the 0- and 1-mode satisfies $\Delta_{01}({\bf K}, j)
\equiv \omega_0({\bf K}, j)- \omega_1({\bf K}, j) = \sqrt{3 \bar{\jmath}}
(1-\sqrt{j})$. I.e. for $j=1$, the optical and acoustic mode display a touching
point at this momentum.

\begin{thebibliography}{99}



\bibitem{Anderson1973}
P. W. Anderson, Mater. Res. Bull. {\bf 8}, 153160 (1973).

\bibitem{Fazekas1974}
P. Fazekas and P. W. Anderson, Phil. Mag. {\bf 30}, 423 (1974).

\bibitem{Misguich2005}
G. Misguich and C. Lhuillier, in Frustrated Spin Systems (World Scientific,
2005), pp. 229-306.

\bibitem{Balents2010}
L. Balents, Nature \textbf{464}, 199-208 (2010).

\bibitem{Savary2017}
L. Savary and L. Balents, Rep. Prog. Phys. \textbf{80}, 016502 (2016).

\bibitem{Knolle2019}
J. Knolle and R. Moessner,
Annu. Rev. Condens. Matter Phys. {\bf 10}, 451 (2019).

\bibitem{Wen2002}
X.-G. Wen, Phys. Rev. B {\bf 65}, 165113 (2002).

\bibitem{Henelius2000}
P. Henelius and A. W. Sandvik, Phys. Rev. B {\bf 62}, 1102 (2000).

\bibitem{Kitaev2006}
A. Kitaev, Ann. Phys. (NY) \textbf{321}, 2 (2006).

\bibitem{Nasu2014}
J. Nasu, M. Udagawa, and Y. Motome, Phys. Rev. Lett. {\bf 113}, 197205 (2014).

\bibitem{Balents2002}
L. Balents, M. P. A. Fisher, and S. M. Girvin,
Phys. Rev. B {\bf 65}, 224412 (2002).

\bibitem{Rokhsar1988}
D. S. Rokhsar and S. A. Kivelson,
Phys. Rev. Lett. {\bf 61}, 2376 (1988).

\bibitem{Moessner2001}
R. Moessner and S. L. Sondhi,
Phys. Rev. Lett. {\bf 86}, 1881 (2001).

\bibitem{Moessner2001a}
R. Moessner, S. L. Sondhi, and E. Fradkin,
Phys. Rev. B {\bf 65}, 024504 (2001).

\bibitem{Sheng2005}
D. N. Sheng and L. Balents,
Phys. Rev. Lett. {\bf 94}, 146805 (2005).

\bibitem{Isakov2006a}
S. V. Isakov, Y. B. Kim, and A. Paramekanti,
Phys. Rev. Lett. {\bf 97}, 207204 (2006).

\bibitem{Isakov2007}
S. V. Isakov, A. Paramekanti, and Y. B. Kim,
Phys. Rev. B {\bf 76}, 224431 (2007).

\bibitem{Dang2011}
L. Dang, S. Inglis, and R. G. Melko,
Phys. Rev. B {\bf 84}, 132409 (2011).

\bibitem{Isakov2011}
S. V. Isakov, M. B. Hastings, and R. G. Melko,
Nature Phys. {\bf 7}, 772 (2011).

\bibitem{Isakov2012}
S. V. Isakov, R. G. Melko, and M. B. Hastings,
Science {\bf 335}, 193 (2012).

\bibitem{Wegner1971}
F. J. Wegner, Journal of Mathematical Physics {\bf 12}, 2259 (1971).

\bibitem{Wen1991}
X. G. Wen, Phys. Rev. B {\bf 44}, 2664 (1991).

\bibitem{Kitaev2003}
A. Yu. Kitaev, Annals of Phys. {\bf 303}, 2 (2003).

\bibitem{Sachdev2018}
S. Sachdev,
Rep. Prog. Phys. {\bf 82}, 014001 (2018).

\bibitem{Wang2017}
Y.-C. Wang, C. Fang, M. Cheng, Y. Qi, and Z. Y. Meng,
arXiv:1701.01552.

\bibitem{Becker2018}
J. Becker and S. Wessel, Phys. Rev. Lett. {\bf 121}, 077202 (2018).

\bibitem{Becker2019}
J. Becker and S. Wessel, Phys. Rev. B {\bf 100}, 241113(R) (2019).

\bibitem{Chubukov1994a}
A. V. Chubukov, T. Senthil, and S. Sachdev,
Phys. Rev. Lett. {\bf 72}, 2089 (1994).

\bibitem{Chubukov1994b}
A. V. Chubukov, S. Sachdev, and T. Senthil,
Nuclear Physics B {\bf 426}, 601 (1994).

\bibitem{Senthil2002}
T. Senthil and O. Motrunich,
Phys. Rev. B {\bf 66}, 205104 (2002).

\bibitem{Isakov2005}
S. V. Isakov, T. Senthil, and Y. B. Kim,
Phys. Rev. B {\bf 72}, 174417 (2005).

\bibitem{Ballesteros1996}
H. G. Ballesteros, L. A. Fern\'{a}ndez, V. Mart\'{i}n-Mayor,
and A. Mu\~{n}oz Sudupe, Physics Letters B {\bf 387}, 125 (1996).

\bibitem{Calabrese2002}
P. Calabrese, A. Pelissetto, and E. Vicari,
Phys. Rev. E {\bf 65}, 046115 (2002).

\bibitem{Campostrini2001}
M. Campostrini, M. Hasenbusch, A. Pelissetto, P. Rossi, and E. Vicari,
Phys. Rev. B {\bf 63}, 214503 (2001).


\bibitem{Mila1998}
F. Mila, Phys. Rev. Lett. {\bf 81}, 2356 (1998).

\bibitem{Mambrini2000}
M. Mambrini and F. Mila, Eur. Phys. J. B {\bf 17}, 651 (2000).

\bibitem{Zhitomirsky2005}
M. E. Zhitomirsky, Phys. Rev. B {\bf 71}, 214413 (2005).

\bibitem{Schaffer2017}
R. Schaffer, Y. Huh, K. Hwang, and Y. B. Kim,
Phys. Rev. B {\bf 95}, 054410 (2017).

\bibitem{Pollmann2017}
C. Repellin, Y.-C. He, and F. Pollmann, Phys. Rev. B {\bf 96}, 205124 (2017).

\bibitem{Iqbal2018}
Y. Iqbal, D. Poilblanc, R. Thomale, and F. Becca,
Phys. Rev. B {\bf 97}, 115127 (2018).

\bibitem{Iqbal2019}
M. Iqbal, D. Poilblanc, and N. Schuch, Phys. Rev. B {\bf 101}, 155141 (2020).


\bibitem{Santos2004}
L. Santos, M. A. Baranov, J. I. Cirac, H.-U. Everts, H. Fehrmann,
and M. Lewenstein, Phys. Rev. Lett. {\bf 93}, 030601 (2004).

\bibitem{Damski2005}
B.~Damski, H.~Fehrmann, H.-U.~Everts, M.~Baranov, L.~Santos, and M.~Lewenstein,
Phys. Rev. A {\bf 72}, 053612 (2005).

\bibitem{Windpassinger2013}
P. Windpassinger and K. Sengstock,
Rep. Prog. Phys. {\bf 76}, 086401 (2013).

\bibitem{Jo2012}
G.-B. Jo, J. Guzman, C. K. Thomas, P. Hosur, A. Vishwanath,
and D. M. Stamper-Kurn, Phys. Rev. Lett. {\bf 108}, 045305 (2012).

\bibitem{Barter2020}
T.~H.~Barter, T.-H.~Leung, M.~Okano, M.~Block, N.~Y.~Yao, and
D.~M.~Stamper-Kurn, Phys. Rev. A {\bf 101}, 011601(R) (2020).


\bibitem{Sandvik1992}
A. W. Sandvik, J. Phys. A {\bf 25}, 3667 (1992).

\bibitem{Sandvik1999}
A. W. Sandvik, Phys. Rev. B {\bf 59}, R14157 (1999).

\bibitem{Syljuasen2002}
O. F. Sylju\aa{}sen and A. W. Sandvik, Phys. Rev. E {\bf 66}, 046701 (2002).

\bibitem{Melko2005}
R. G. Melko and A. W. Sandvik,
Phys. Rev. E {\bf 72}, 026702 (2005).


\bibitem{Cuccoli2003}
A.~Cuccoli, T.~Roscilde, V.~Tognetti, R.~Vaia, and P.~Verrucchi,
Phys. Rev. B {\bf 67}, 104414 (2003).

\bibitem{Sandvik1997}
A. W. Sandvik, Phys. Rev. B {\bf 56}, 11678 (1997).

\bibitem{Pollock1987}
E. L. Pollock and D. M. Ceperley,
Phys. Rev. B {\bf 36}, 8343 (1987).

\bibitem{Harada1998}
K. Harada and N. Kawashima,
J. Phys. Soc. Jpn. {\bf 67}, 2768 (1998).


\bibitem{Skilling1984}
J. Skilling and R. K. Bryan, Mon. Not. R. Astron. Soc. {\bf 211}, 111 (1984).

\bibitem{Bryan1990}
R. Bryan, Eur. Biophys. J. {\bf 18}, 165174 (1990).

\bibitem{Jarrell1996}
M. Jarrell and J. Gubernatis, Phys. Rep. {\bf 269}, 133 (1996).


\bibitem{Fisher1989}
M.~P.~A.~Fisher, P.~B.~Weichman, G.~Grinstein, and D.~S.~Fisher,
Phys. Rev. B {\bf 40}, 546 (1989).

\bibitem{NoteDang2011} In \cite{Dang2011} $J$ is rescaled by a factor of $2$,
as compared to this work, resulting in $(K/J)_c\approx 21.8$.

\end{thebibliography}
\end{document}